\documentclass[reprint,showpacs,preprintnumbers, amsmath,amssymb,aps,prc,]{revtex4-1}
\usepackage{amsfonts}
\usepackage{amsmath}
\usepackage{amssymb}
\usepackage{graphicx,color}
\usepackage{rotating}
\usepackage{dcolumn}
\usepackage{bm,ulem}
\usepackage[colorlinks=true, pdfstartview=FitV, linkcolor=blue, 
citecolor=blue,urlcolor=navyblue]{hyperref}

\begin{document}

\title{Critical parameters of consistent relativistic mean-field models} 

\author{O. Louren\c{c}o$^1$, M. Dutra$^2$, and D. P. Menezes$^3$} 

\affiliation{$^1$Universidade Federal do Rio de Janeiro, 27930-560, Maca\'e, RJ, Brazil \\
$^2$Departamento de Ci\^encias da Natureza - IHS, Universidade Federal
Fluminense, 28895-532 Rio das Ostras, RJ, Brazil \\
$^3$Depto de F\'{\i}sica - CFM - Universidade Federal de Santa Catarina, Florian\'opolis 
- SC - CP. 476 - CEP 88.040 - 900 - Brazil}
\date{\today}

\begin{abstract}
 In the present work, the critical temperature, critical pressure and
 critical density, known as the critical parameters related to the liquid-gas phase transition are
calculated for 34 relativistic mean-field models, which were shown to satisfy
nuclear matter constraints in a comprehensive study involving 263
models. The compressibility factor was calculated and all 34 models
present values lower than the one obtained with the van der Waals
equation of state. The critical temperatures were compared with
experimental data and just two classes of models can reach values
close to them. A correlation between the critical parameters and the
incompressibility was obtained.

\end{abstract}


\maketitle

\section{Introduction}

The understanding of nuclear matter properties is of fundamental
importance as a guide towards more specific subjects, such as nuclear
and hadron spectroscopy, heavy-ion collisions, caloric curves and
negative heat capacities, nuclear multifragmentation and distillation
effects, neutron stars and the possible existence of the pasta phase
in its core and even the QCD phase diagram and its phase transitions.
 At low densities and relatively low temperatures (below 20 MeV), 
nuclear matter can evolve through different phase separation 
boundaries and the construction of binodals
depicts very well this problem. Another important aspect is the
investigation of instability boundaries and the spinodals are 
used to separate unstable from stable matter. These sections (binodals
and spinodals) are just a reflex of the well known fact that at low
densities, nuclear matter undergoes a first order phase transition,
which belongs to the liquid-gas universality class \cite{lg,vdw1,chomaz}.

A seminal work on the use
of relativistic models to describe multicomponent systems (in nuclear
matter, the components are protons and neutrons) is
\cite{vdw1}. This extremely didactic paper clearly shows how the
geometrical Maxwell construction can be used to determine the amount of
particles (proton fraction) and the related chemical potentials in the
coexistence phase and the construction of the binodal section. 
As far as unstable matter is concerned, the instabilities a system may
present are related to the possible phase transitions it can undertake
\cite{chomaz}.  Spinodal sections are obtained from the
derivative of the free energy of the system with respect to the
chemical potentials of its components.
The spinodal instability is known to lead to a
liquid-gas phase transition with the restoration of the isospin
symmetry at a certain density. 

In Ref.~\cite{vdw1}, a three dimensional plot (see Fig. 7) shows the
phase coexistence boundary in a pressure-temperature-proton fraction
plane, from where it is seen that the critical temperature always takes place in
the symmetric matter. Analogously, in \cite{avancini2006}, it was
shown that the instability region
 decreases with the increase of the temperature up to a certain {\it
   critical temperature}, which is related to a {\it critical
   pressure} and {\it critical density}. For temperatures larger than
 the critical temperature, the system is stable. Once again, these
 {\it critical parameters} always take place at proton fraction $0.5$,
 i.e., symmetric nuclear matter (see Table IV).

Nevertheless, the values of these critical parameters are model
dependent and there are many nonrelativistic~\cite{gulminelli2012}
and relativistic models \cite{avancini2006} in the market, which 
can be used to calculate binodals and spinodals. The references just
given show only a few of them. In the present work we restrict our
investigation to specific relativistic mean-field (RMF) models, which were
shown to satisfy important nuclear matter bulk properties in
Ref.~\cite{rmf}.  They are named here as
consistent relativistic mean field parametrizations (CRMF) 
and in the next section a more detailed explanation on this
choice is made. 

After the presentation of 34 CRMF models, we show how the critical parameters
are obtained and their values are displayed and compared with
experimental results. The conclusions are drawn in the last section of
the present work.

\section{Consistent relativistic mean-field models}

The analysis performed in Ref.~\cite{rmf} pointed out to only $35$
parametrizations, out of 
$263$ investigated, simultaneously approved in seven distinct nuclear matter constraints. 
These consistent RMF parametrizations had their bulk and thermodynamical quantities 
compared to respective theoretical/experimental data from symmetric nuclear matter (SNM), 
pure neutron matter (PNM) and a mixture of both, namely, symmetry energy and its slope 
evaluated at the saturation density $\rho_0$, and the ratio of the symmetry energy at 
$\rho_0/2$ to its value at $\rho_0$ (MIX). These detailed constraints are specified in 
Table~\ref{set2a}.
\begin{table}[hb!]
\scriptsize
\begin{ruledtabular}
\caption{Set of updated constraints (SET2a) used in Ref.~\cite{rmf}. See that reference 
for more details concerning each constraint.}
\centering
\begin{tabular}{lccc}
Constraint & Quantity      & Density Region       & Range of constraint \\ 
\hline
SM1    & $K_0$     & at $\rho_0$  & 190 $-$ 270 MeV \\
SM3a   & $P(\rho)$ & $2<\frac{\rho}{\rho_0}<5$ & Band Region \\
SM4    & $P(\rho)$ & $1.2<\frac{\rho}{\rho_0}<2.2$    & Band Region\\
PNM1   & $\mathcal{E}_{\mbox{\tiny PNM}}/\rho$ & $0.017<\frac{\rho}{\rho_{\rm
o}}<0.108$   & Band Region \\
MIX1a  & $J$       & at $\rho_0$  & 25 $-$ 35 MeV \\
MIX2a  & $L_0$     & at $\rho_0$  & 25 $-$ 115 MeV \\
MIX4   & $\frac{\mathcal{S}(\rho_0/2)}{J}$  & at $\rho_0$ and $\rho_0/2$ & 0.57 $-$ 0.86\\
\end{tabular}
\label{set2a}
\end{ruledtabular}
\end{table}

In Ref. \cite{rmf}, the models were divided into seven different
categories and only three of these categories included models that satisfies the
imposed constraints.
Among the $35$ CRMF parametrizations, $30$ of them are of type $4$~\cite{rmf}, 
i.e., the Lagrangian density comprises nonlinear $\sigma$ and $\omega$
terms and cross terms involving these fields. They 
are: BKA20~\cite{bka}, BKA22~\cite{bka}, BKA24~\cite{bka}, BSR8~\cite{bsr}, 
BSR9~\cite{bsr}, BSR10~\cite{bsr}, BSR11~\cite{bsr}, BSR12~\cite{bsr}, BSR15~\cite{bsr}, 
BSR16~\cite{bsr}, BSR17~\cite{bsr}, BSR18~\cite{bsr}, BSR19~\cite{bsr}, BSR20~\cite{bsr}, 
FSU-III~\cite{fsu-iii-iv}, FSU-IV~\cite{fsu-iii-iv}, FSUGold~\cite{fsugold}, 
FSUGold4~\cite{fsugold4}, FSUGZ03~\cite{fsugz}, FSUGZ06~\cite{fsugz}, G2*~\cite{g2*}, 
IU-FSU~\cite{iufsu}, Z271s2~\cite{z271}, Z271s3~\cite{z271}, Z271s4~\cite{z271}, 
Z271s5~\cite{z271}, Z271s6~\cite{z271}, Z271v4~\cite{z271}, Z271v5~\cite{z271}, and 
Z271v6~\cite{z271}. The Lagrangian density that describes such parametrizations is
\begin{eqnarray}
\mathcal{L}_{\rm NL} &=& \overline{\psi}(i\gamma^\mu\partial_\mu - M)\psi 
+ g_\sigma\sigma\overline{\psi}\psi - g_\omega\overline{\psi}\gamma^\mu\omega_\mu\psi 
\nonumber \\
&-& \frac{g_\rho}{2}\overline{\psi}\gamma^\mu\vec{\rho}_\mu\vec{\tau}\psi
+ \frac{1}{2}(\partial^\mu \sigma \partial_\mu \sigma 
- m^2_\sigma\sigma^2) - \frac{A}{3}\sigma^3 
\nonumber\\
&-&  \frac{B}{4}\sigma^4 -\frac{1}{4}F^{\mu\nu}F_{\mu\nu} 
+ \frac{1}{2}m^2_\omega\omega_\mu\omega^\mu + 
\frac{C}{4}(g_\omega^2\omega_\mu\omega^\mu)^2 
\nonumber \\
&-& \frac{1}{4}\vec{B}^{\mu\nu}\vec{B}_{\mu\nu} + 
\frac{1}{2}m^2_\rho\vec{\rho}_\mu\vec{\rho}^\mu
+ \frac{1}{2}{\alpha_3'}g_\omega^2
g_\rho^2\omega_\mu\omega^\mu\vec{\rho}_\mu\vec{\rho}^\mu
\nonumber\\
&+& g_\sigma g_\omega^2\sigma\omega_\mu\omega^\mu
\left(\alpha_1+\frac{1}{2}{\alpha_1'}g_\sigma\sigma\right)
\nonumber\\
&+& g_\sigma g_\rho^2\sigma\vec{\rho}_\mu\vec{\rho}^\mu
\left(\alpha_2+\frac{1}{2}{\alpha_2'}g_\sigma\sigma\right),
\label{lomegarho}
\end{eqnarray}
with $F_{\mu\nu}=\partial_\nu\omega_\mu-\partial_\mu\omega_\nu$
and $\vec{B}_{\mu\nu}=\partial_\nu\vec{\rho}_\mu-\partial_\mu\vec{\rho}_\nu$. The nucleon 
mass is $M$ and the meson masses are $m_j$, for $j=\sigma,\omega,$ and $\rho$.

The other $4$ CRMF approved parametrizations present density dependent (DD) coupling 
constants. Two of them are standard DD parametrizations: \mbox{DD-F}~\cite{ddf} and 
TW99~\cite{tw99}, and the remaining two also present the $\delta$ meson in their 
structures: \mbox{DDH$\delta$}~\cite{ddhd} and \mbox{DD-ME$\delta$}~\cite{ddmed}. The 
Lagrangian density of all of them is expressed as,
\begin{eqnarray}
\mathcal{L}_{\rm DD} &=& \overline{\psi}(i\gamma^\mu\partial_\mu - M)\psi 
+ \Gamma_\sigma(\rho)\sigma\overline{\psi}\psi 
- \Gamma_\omega(\rho)\overline{\psi}\gamma^\mu\omega_\mu\psi 
\nonumber\\
&-&\frac{\Gamma_\rho(\rho)}{2}\overline{\psi}\gamma^\mu\vec{\rho}_\mu\vec{\tau}
\psi + \Gamma_\delta(\rho)\overline{\psi}\vec{\delta}\vec{\tau}\psi 
- \frac{1}{4}F^{\mu\nu}F_{\mu\nu}
\nonumber \\
&+& \frac{1}{2}(\partial^\mu \sigma \partial_\mu \sigma - m^2_\sigma\sigma^2)
 + \frac{1}{2}m^2_\omega\omega_\mu\omega^\mu 
-\frac{1}{4}\vec{B}^{\mu\nu}\vec{B}_{\mu\nu}
\nonumber \\
&+& \frac{1}{2}m^2_\rho\vec{\rho}_\mu\vec{\rho}^\mu + 
\frac{1}{2}(\partial^\mu\vec{\delta}\partial_\mu\vec{\delta} 
- m^2_\delta\vec{\delta}^2),
\label{dldd}
\end{eqnarray}
where
\begin{eqnarray}
\Gamma_i(\rho) &=& \Gamma_i(\rho_0)f_i(x);\quad
f_i(x) = a_i\frac{1+b_i(x+d_i)^2}{1+c_i(x+d_i)^2},
\label{gamadefault}
\end{eqnarray}
for $i=\sigma,\omega$, and $x=\rho/\rho_0$. The Lagrangian density describing the 
\mbox{DD-F} and TW99~\cite{tw99} parametrizations is the same as the one in Eq.~(\ref{dldd}) 
when the meson $\delta$ is not taken into account.

The last CRMF parametrization is a point-coupling 
model~\cite{pc1,pc2,pc3,pc4,pc5,pc6,pc7}: FA3~\cite{fa3}. Here, we do not investigate 
such model since in a previous work~\cite{stars} we have showed it is not capable of 
generating, already in the  zero temperature regime, a mass radius curve for neutron stars, 
due to a very particular behavior in the high-density regime, namely, a fall in the 
pressure versus energy density ($\varepsilon$) curve near $\varepsilon = 4.1$~fm$^{-4}$. 
For that reason, we have decided to discard this particular parametrization from our finite temperature 
analysis.

\section{Results from finite temperature regime}

We next present only the main formulae for the calculation of the
critical parameters. All other calculations and the complete equations
of state are given in detail in Ref.~\cite{rmf} and we do reproduce them
here. 

\subsection{Critical parameters, and model dependence in the liquid phase}

The CRMF critical parameters are obtained directly from the
thermodynamical pressure ($P$) of these models once the  
following conditions in the $P\times\rho$ plane are imposed:
\begin{eqnarray}
P_c=P(\rho_c , T_c),\quad 
\frac{\partial P}{\partial\rho}\bigg|_{\rho_c , T_c}=0,\quad
\frac{\partial^2 P}{\partial\rho^2}\bigg|_{\rho_c , T_c}=0,\quad
\label{conditions}
\end{eqnarray}
where $P_c$, $\rho_c$ and $T_c$ are, respectively, the critical pressure, density and 
temperature. These three critical parameters define a unique critical point.
Such constraints can be used because hadronic mean-field models 
present the same features exhibited by the van der Waals model, i. e., a liquid gas phase 
transition at temperatures smaller than $T_c$, see 
Refs.~\cite{vdw1,vdw2,vdw3,vdw4,vdw5,vdw6,wu,dhiman,lenske}, for instance.

From the Lagrangian density in Eq.~(\ref{lomegarho}), one can derive the expression for 
symmetric nuclear matter ($\gamma=4$) pressure, by following, for example, the steps 
indicating in Ref.~\cite{serot}. The result is 
\begin{align}
P_{\rm NL} &= - \frac{1}{2}m^2_\sigma\sigma^2 - \frac{A}{3}\sigma^3 -
\frac{B}{4}\sigma^4 + \frac{1}{2}m^2_\omega\omega_0^2 
+ \frac{C}{4}(g_\omega^2\omega_0^2)^2
\nonumber\\
+& g_\sigma g_\omega^2\sigma\omega_0^2
\left(\alpha_1+\frac{1}{2}{\alpha_1'}g_\sigma\sigma\right) 
\nonumber\\
+& \dfrac{\gamma}{6\pi^2}\int_0^{\infty}\dfrac{dk\,k^4}{(k^2 + 
{M^*}^2)^{1/2}}\left[n(k,T,\mu^*)+\bar{n}(k,T,\mu^*)\right],\nonumber\\
\label{pnl}
\end{align}
where
\begin{align}
n(k,T,\mu^*) &= \frac{1}{e^{(E^*-\mu^*)/T}+1},\quad\mbox{and}\nonumber\\
\bar{n}(k,T,\mu^*) &= \frac{1}{e^{(E^*+\mu^*)/T}+1}
\label{fermi-dirac}
\end{align}
are the Fermi-Dirac distributions for particles and antiparticles, respectively. The 
effective energy, nucleon mass, and chemical potential are $E^*=(k^2+{M^*}^2)^{1/2}$, 
$M^*=M-g_\sigma\sigma$, and $\mu^*=\mu-g_\omega\omega_0$, respectively. Furthermore, the 
(classical) mean-field values of $\sigma$ and $\omega_0$ are found by solving the 
following system of equations,
\begin{eqnarray}
m^2_\sigma\sigma &=& g_\sigma\rho_s - A\sigma^2 - B\sigma^3 
+g_\sigma g_\omega^2\omega_0^2(\alpha_1+{\alpha_1'}g_\sigma\sigma)
\\
m_\omega^2\omega_0 &=& g_\omega\rho - Cg_\omega(g_\omega \omega_0)^3 
- g_\sigma g_\omega^2\sigma\omega_0(2\alpha_1+{\alpha_1'}g_\sigma\sigma),
\nonumber\\
\end{eqnarray}
with
\begin{align}
\rho&=\dfrac{\gamma}{2\pi^2}\int_0^{\infty}dk\,k^2
\left[n(k,T,\mu^*)-\bar{n}(k,T,\mu^*)\right],
\label{rho}
\\
\rho_s &= \dfrac{\gamma}{2\pi^2}\int_0^{\infty}\frac{dk\,M^*k^2}{(k^2+{M^*}^2)^{1/2}}
\left[n(k,T,\mu^*)+\bar{n}(k,T,\mu^*)\right].
\label{rhos}
\end{align}

It is worth noticing in these derivations that $\left<\vec{\rho}_\mu\right>\equiv 
\bar{\rho}_{0(3)}$, and $\left<\vec{\delta}\right>\equiv\delta_{(3)}$ are vanishing, since 
we are restricted to the symmetric nuclear matter system, in which $\rho_p=\rho_n$ and 
${\rho_s}_p={\rho_s}_n$. For that reason, terms in Eq.~(\ref{lomegarho}) involving 
specifically these fields do not contribute to the thermodynamical quantities of the 
model, or in any other calculations in the mean-field approximation.

The same procedures exposed in Ref.~\cite{serot} are also used in order to generate the 
pressure for the density dependent model described by
Eq.~(\ref{dldd}). Once again, the fields $\bar{\rho}_{0(3)}$ and $\delta_{(3)}$ 
do not contribute for the calculations. Therefore, the thermodynamics of \mbox{DD-F} and 
TW99 parametrizations is exactly the same of the \mbox{DDH$\delta$} and 
\mbox{DD-ME$\delta$} ones. In particular, the symmetric nuclear matter pressure reads
\begin{align}
P_{\rm DD} &= \rho\Sigma_R(\rho)- \frac{1}{2}m^2_\sigma\sigma^2 +
\frac{1}{2}m^2_\omega\omega_0^2 
\nonumber\\
+& \frac{\gamma}{6\pi^2}\int_0^{\infty}\dfrac{dk\,k^4}{(k^2 + 
{M^*}^2)^{1/2}}\left[n(k,T,\mu^*)+\bar{n}(k,T,\mu^*)\right],
\nonumber\\
\label{pressuredd}
\end{align}
with the rearrangement term defined as 
\begin{eqnarray}
\Sigma_R(\rho)=\frac{\partial\Gamma_\omega}{\partial\rho}\omega_0\rho
-\frac{\partial\Gamma_\sigma}{\partial\rho}\sigma\rho_s.
\end{eqnarray}
The mean-fields $\sigma$ and $\omega_0$ are given by
\begin{eqnarray}
\sigma = \frac{\Gamma_\sigma(\rho)}{m_\sigma^2}\rho_s,\quad\mbox{and}\quad
\omega_0 = \frac{\Gamma_\omega(\rho)}{m_\omega^2}\rho,
\label{mfdd}
\end{eqnarray}
with the functional forms of $\rho$ and $\rho_s$ given as in the nonlinear model, 
Eqs.~(\ref{rho})-(\ref{rhos}), with the same distributions functions of 
Eq.~(\ref{fermi-dirac}), and the same form for the effective energy $E^*$. The effective 
nucleon mass and chemical potential are now given, respectively, by 
$M^*=M-\Gamma_\sigma(\rho)\sigma$, and $\mu^*=\mu-\Gamma_\omega(\rho)\omega_0 - 
\Sigma_R(\rho)$.

Since the expressions given in Eqs.~(\ref{pnl}) and (\ref{pressuredd}) are completely 
determined, we are able to apply the conditions to calculate the
critical point given in Eq.~(\ref{conditions}) and then obtain 
$P_c$, $\rho_c$ and $T_c$ for each of the CRMF parametrizations. These 
results are presented in Table~\ref{tabcritical}. Also in this Table, we furnish the 
compressibility factor, defined as $Z_c = P_c/\rho_cT_c$. For the van der Waals (vdW) 
equation of state (EOS), for example, this quantity has a value of $0.375$, independently 
of the fluid described by it. This is a direct consequence of the
universality of the vdW EOS. We have divided the 34 CRMF models into 6
families. Notice that all CRMF parametrizations present $Z_c<0.375$.
\begin{table}[!htb]
\scriptsize
\caption{Critical parameters ($T_c$, $\rho_c$ and $P_c$), and compressibility factor 
($Z_c = P_c/\rho_cT_c$) of CRMF parametrizations.}
\centering
\begin{tabular}{l|c|c|c|c|c}
\hline\hline
Model & $T_{\rm c}$ (MeV)  & $\rho_c$ (fm$^{-3}$)  & $P_c$~(MeV/fm$^3$)  & 
~~$\dfrac{\rho_c}{\rho_0}$~~  & $Z_c$ \\
\hline
BKA20   & $14.92$ & $0.0458$ & $0.209$ & $0.314$ & $0.306$ \\ 
BKA22   & $13.91$ & $0.0442$ & $0.178$ & $0.300$ & $0.290$ \\ 
BKA24   & $13.83$ & $0.0450$ & $0.177$ & $0.306$ & $0.284$ \\ 
\hline
BSR8    & $14.17$ & $0.0440$ & $0.185$ & $0.300$ & $0.297$ \\ 
BSR9    & $14.11$ & $0.0450$ & $0.185$ & $0.305$ & $0.291$ \\
BSR10   & $13.90$ & $0.0439$ & $0.176$ & $0.297$ & $0.288$ \\ 
BSR11   & $14.00$ & $0.0442$ & $0.179$ & $0.301$ & $0.289$ \\ 
BSR12   & $14.15$ & $0.0448$ & $0.185$ & $0.304$ & $0.292$ \\ 
BSR15   & $14.53$ & $0.0456$ & $0.199$ & $0.313$ & $0.300$ \\ 
BSR16   & $14.44$ & $0.0454$ & $0.196$ & $0.311$ & $0.299$ \\ 
BSR17   & $14.32$ & $0.0451$ & $0.191$ & $0.308$ & $0.296$ \\ 
BSR18   & $14.25$ & $0.0451$ & $0.189$ & $0.309$ & $0.294$ \\ 
BSR19   & $14.28$ & $0.0451$ & $0.190$ & $0.308$ & $0.295$ \\ 
BSR20   & $14.41$ & $0.0464$ & $0.197$ & $0.318$ & $0.295$ \\
\hline 
FSU-III & $14.75$ & $0.0461$ & $0.205$ & $0.311$ & $0.301$ \\ 
FSU-IV  & $14.75$ & $0.0461$ & $0.205$ & $0.311$ & $0.301$ \\ 
FSUGold & $14.75$ & $0.0461$ & $0.205$ & $0.311$ & $0.301$ \\ 
FSUGold4& $14.80$ & $0.0456$ & $0.204$ & $0.309$ & $0.302$ \\ 
FSUGZ03 & $14.11$ & $0.0450$ & $0.185$ & $0.305$ & $0.291$ \\ 
FSUGZ06 & $14.44$ & $0.0454$ & $0.196$ & $0.311$ & $0.299$ \\ 
IU-FSU  & $14.49$ & $0.0457$ & $0.196$ & $0.295$ & $0.296$ \\ 
\hline
G2*     & $14.38$ & $0.0468$ & $0.192$ & $0.305$ & $0.285$ \\ 
\hline
Z271s2  & $17.97$ & $0.0509$ & $0.303$ & $0.343$ & $0.331$ \\ 
Z271s3  & $17.97$ & $0.0509$ & $0.303$ & $0.343$ & $0.331$ \\ 
Z271s4  & $17.97$ & $0.0509$ & $0.303$ & $0.343$ & $0.331$ \\ 
Z271s5  & $17.97$ & $0.0509$ & $0.303$ & $0.343$ & $0.331$ \\ 
Z271s6  & $17.97$ & $0.0509$ & $0.303$ & $0.343$ & $0.331$ \\ 
Z271v4  & $17.97$ & $0.0509$ & $0.303$ & $0.343$ & $0.331$ \\ 
Z271v5  & $17.97$ & $0.0509$ & $0.303$ & $0.343$ & $0.331$ \\ 
Z271v6  & $17.97$ & $0.0509$ & $0.303$ & $0.343$ & $0.331$ \\ 
\hline
DD-F    & $15.24$ & $0.0505$ & $0.245$ & $0.343$ & $0.318$ \\ 
TW99    & $15.17$ & $0.0509$ & $0.241$ & $0.332$ & $0.312$ \\ 
DDH$\delta$   & $15.17$ & $0.0509$ & $0.241$ & $0.332$ & $0.312$ \\ 
DD-ME$\delta$ & $15.32$ & $0.0491$ & $0.235$ & $0.323$ & $0.312$
\\ 
\hline \hline
\end{tabular}
\label{tabcritical}
\end{table}

\begin{figure}[!htb]
\centering
\includegraphics[scale=0.35]{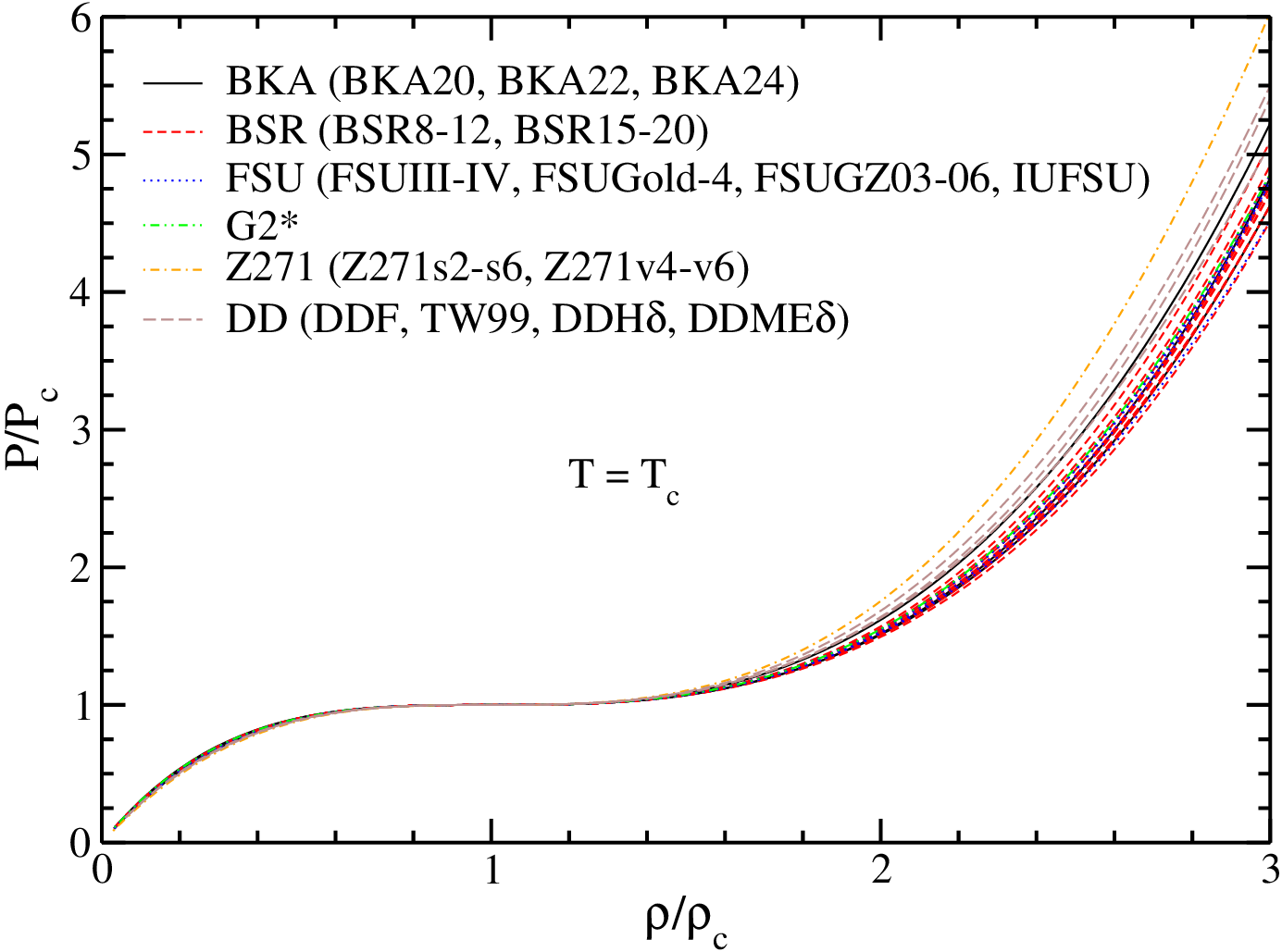}
\vspace{-0.2cm}
\caption{Pressure as a function of density, both in units of their respective 
critical values, for the CRMF parametrizations at $T=T_c$.} 
\label{pc}
\end{figure}

In Fig.~\ref{pc}, we present the density dependence of the pressure for the CRMF 
parameterizations, in units of $P_c$ and $\rho_c$, all of them at $T=T_c$. 
In this figure, we notice an interesting feature also reported for the Boguta-Bodmer 
models analyzed in Ref.~\cite{vdw5} (see Fig.~1 of this reference), namely, the scaled 
curves are indistinguishable in the gaseous phase ($\rho/\rho_c<1$) and distinct from each 
other, i.e., model dependent, in the liquid phase region ($\rho/\rho_c>1$). The authors 
of Ref.~\cite{vdw5} claimed that in the latter region, the nucleons are confined to a 
smaller phase space, approaching each other progressively and allowing the interactions to 
take place more substantially. This phenomenology is reflected by the scaling, exhibited 
for $\rho/\rho_c<1$, and absent in the remaining region. As the structure of the 
parametrizations analyzed in Ref.~\cite{vdw5} was restricted to RMF models presenting only 
third- and forth order self-interactions in the scalar field $\sigma$ (Boguta-Bodmer 
model), it was difficult to generalize such result to any RMF parametrization. However, 
here we investigate more sophisticated RMF models, including that one where the couplings 
are density dependent, and the phenomenology of the liquid phase presented in the 
Boguta-Bodmer model was showed again, indicating the general trend of RMF 
parametrizations of any kind in presenting a model dependence in the liquid phase, and a 
scaling in the gaseous one, at symmetric nuclear matter environment.

\subsection{Comparison with experimental data}

\begin{figure}[!htb]
\centering
\includegraphics[scale=0.35]{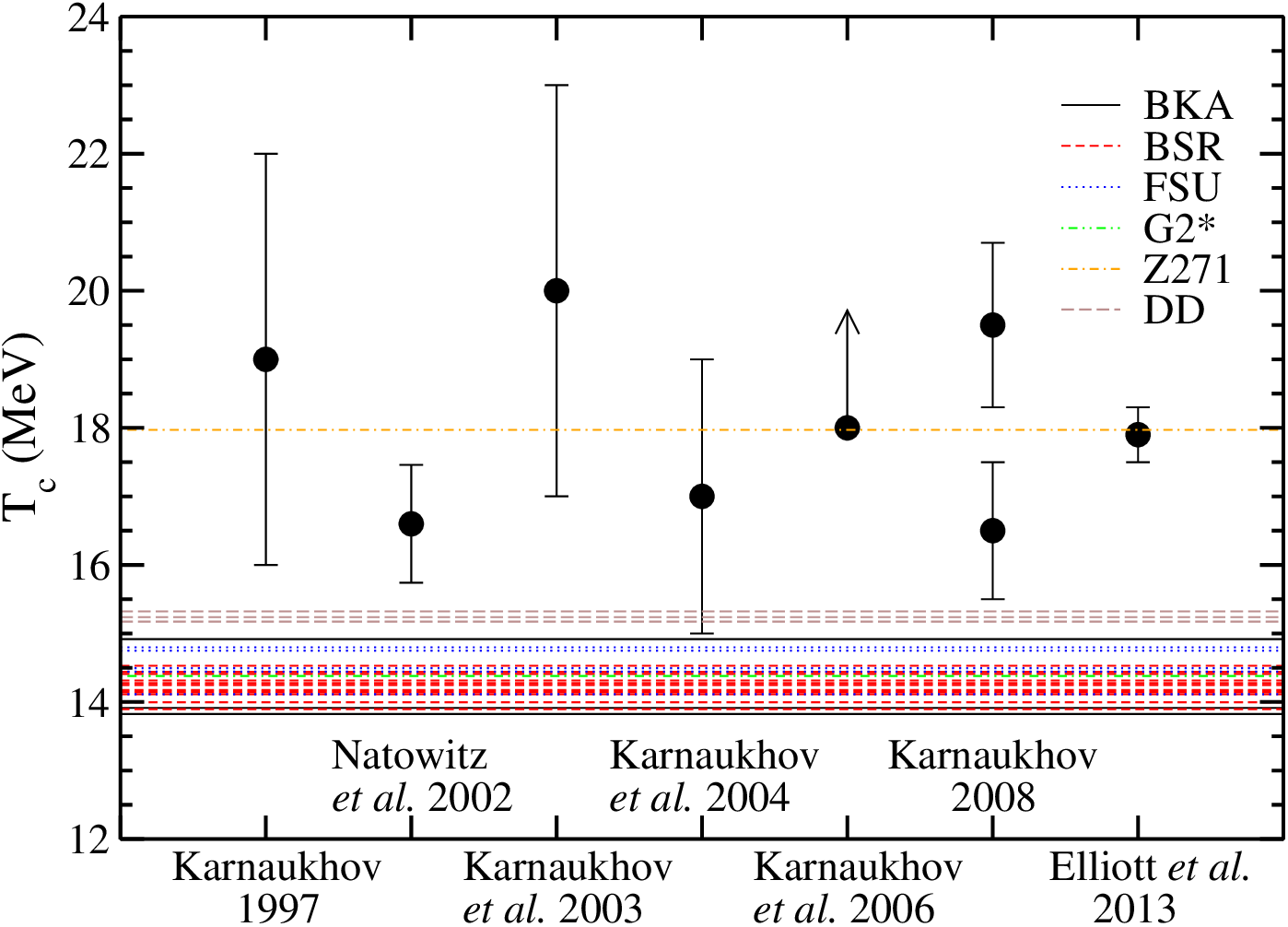}
\vspace{-0.2cm}
\caption{Critical temperature of CRMF parametrizations compared with experimental data 
(circles) collected from the following references: \mbox{Karnaukhov 
1997 \cite{karn1}}, \mbox{Natowitz {\it et al.} 2002 \cite{natowitz}}, \mbox{Karnaukhov 
{\it et al.} 2003 \cite{karn2}}, \mbox{Karnaukhov {\it et al.} 2004 \cite{karn3}}, 
\mbox{Karnaukhov {\it et al.} 2006 \cite{karn4}}, \mbox{Karnaukhov 2008 \cite{karn5}}, 
and \mbox{Elliott {\it et al.} 2013 \cite{elliott}}. The parametrization families are 
indicated as in Fig.~\ref{pc}.} 
\label{tc-exp}
\end{figure}

As a further analysis of the CRMF critical parameters, we compared such 
quantities with known experimental data. Firstly, we compare the critical temperature in 
Fig.~\ref{tc-exp}.

We can see that only a few parametrizations reach some of experimental points. The 
density dependent TW99, \mbox{DD-F}, \mbox{DDH$\delta$} and 
\mbox{DD-ME$\delta$}~\cite{ddmed} present $T_c$ inside the range of $15\leqslant T_c 
\leqslant 19$ MeV~\cite{karn3}, and the family Z271, that encompasses all $8$ related 
parametrizations, has the critical temperature compatible with $5$ of the $8$ experimental 
points, including the more recent one of Ref.~\cite{elliott}.

In this latter work~\cite{elliott}, the authors were able to experimentally determine 
all three critical parameters, unlike previous studies focusing only in $T_c$. For that 
purpose, they have used two types of experiments, namely, compound-nucleus and nuclear 
multifragmentation. In the former, two different nuclei collide with each other and form a 
single compound system, with excitation energy obtained from the energy and masses of the 
subsystems. They have analyzed results from the following compound-nucleus reactions: 
$^{58}$Ni$+^{12}$C$\rightarrow ^{70}$Se and $^{64}$Ni$+^{12}$C$\rightarrow ^{76}$Se, 
performed at the 88-in. cyclotron of the Lawrence Berkeley National 
Laboratory (LBNL)~\cite{lbnl}. In the latter experiment in that study, a beam of 
relativistic incident light particles was used to heat a particular target nucleus. The 
intermediate-mass fragments emitted in this multifragmentation process are essential for 
determination of thermal quantities. In Ref.~\cite{elliott}, the authors also studied 
the multifragmentation reactions, performed by the Indiana Silicon Sphere Collaboration 
at the Alternating Gradient Synchrotron at Brookhaven National Laboratory~\cite{bnl}, and 
by the Equation of State Collaboration at LBNL. The studied reactions were 
$1$~GeV/$c\,\,\,\pi+^{197}$Au, $1$~GeV/nucleon $^{197}$Au$+^{12}$C, $1$~GeV/nucleon 
$^{139}$La$+^{12}$C, and $1$~GeV/nucleon $^{84}$Kr$+^{12}$C. The yields of all these 
reactions were analyzed within a Fisher droplet model, modified to take into account 
asymmetry, Coulomb and finite-size effects, and angular momentum arising from the 
collisions. The analyzed results from all these compound-nucleus and multifragmentation 
reactions, pointed out to $T_c=17.9\pm0.4$~MeV, $P_c=0.31\pm0.07$~MeV/fm$^3$, and 
$\rho_c=0.06\pm0.01$~fm$^{-3}$, for the critical parameters of symmetric nuclear matter.

\begin{figure}[!htb]
\centering
\includegraphics[scale=0.35]{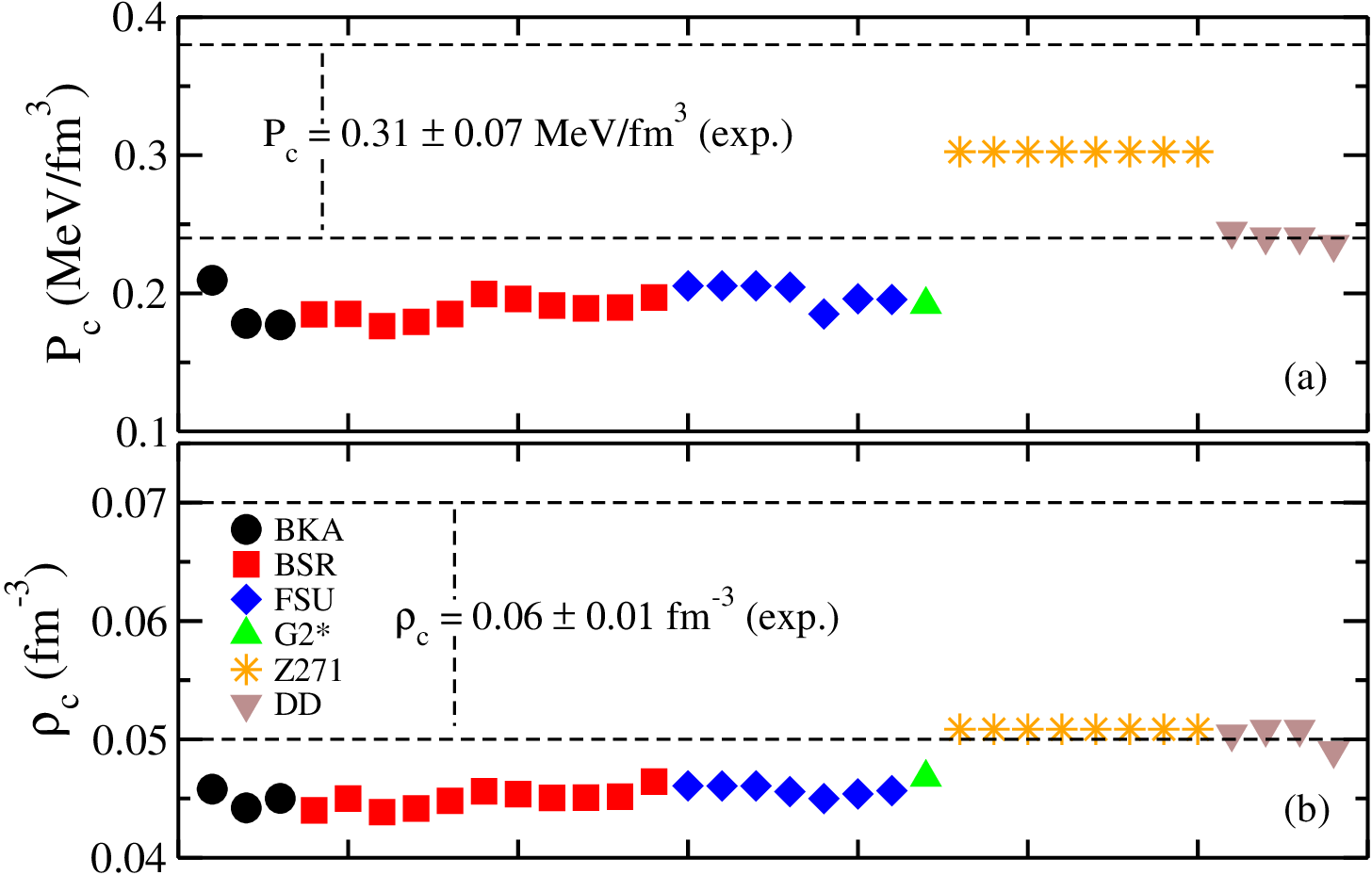}
\vspace{-0.2cm}
\caption{Critical (a) pressure and (b) density for all CRMF parametrizations, compared 
with the corresponding experimental values extracted from Ref.~\cite{elliott}.} 
\label{pc-rhoc-exp}
\end{figure}

As mentioned before, the family of parametrizations named as Z271 has exactly the same 
experimental value of $T_c$ from Ref.~\cite{elliott}, as we can see in Fig.~\ref{tc-exp}. 
For the sake of completeness concerning $P_c$ and $\rho_c$, we have also compared these 
particular critical values of the CRMF parametrizations to those experimental ones of 
Ref.~\cite{elliott}, namely, $P_c=0.31\pm0.07$~MeV/fm$^3$, and 
$\rho_c=0.06\pm0.01$~fm$^{-3}$. The results are depicted in Fig.~\ref{pc-rhoc-exp}. As we 
can see, once more the set of Z271 parametrizations completely agrees with the data. 
Specifically for these critical parameters, we also notice agreement of the density 
dependent model with the experiments. The remaining CRMF parametrizations are not inside 
the boundaries.

By analyzing in detail the Z271 family~\cite{z271}, we observe that in all eight 
parametrizations the couplings $\alpha_1$ and $\alpha_1'$ are vanishing, and $C\ne0$ is 
the only constant that differs these parametrizations from those of the Boguta-Bodmer 
model, see Eq.~(\ref{pnl}). There are no interactions between mesons in this case for 
symmetric nuclear matter, only self-interactions. In some sense, the density dependent 
model has a similar structure, since the nonlinear behavior of the $\sigma$ field can be 
represented somehow in the thermodynamical quantities, by the density dependent constant 
$\Gamma_\sigma(\rho)$. The same occurs with the $\omega_0$ field, i.e., the strength of 
the repulsive interaction is also a density dependent quantity, $\Gamma_\omega(\rho)$. 
Therefore, the DD model can be seen as an effective model in which the nonlinear behavior 
of the scalar and vector fields are included in the density dependence of the respective 
couplings. Such a nonlinear behavior of the fields seems to help the
model in reaching the 
experimental values of the critical parameters of Ref.~\cite{elliott}. In the case of the 
Z271 family, the matching is for all three quantities and for the DD parametrizations, 
only the $T_c$ experimental value is not reached, with the exception of the 
\mbox{DD-ME$\delta$} model in which only $P_c$ data matches. A systematic 
investigation involving a larger number of parametrizations is needed in order to 
definitely confirm our findings. However, the CRMF models strongly suggest such a 
phenomenology.

\subsection{Correlations with the incompressibility}

As a last investigation concerning the critical parameters, we discuss here whether the 
correlations found in Ref.~\cite{rmft} also apply to the CRMF parametrizations. In that 
work, a strong correlation between $T_c$, $P_c$ and $\rho_c$ and the 
incompressibility, $K_0$, obtained at zero temperature regime and at the saturation 
density, was found. For symmetric nuclear matter, the incompressibility of the 
nonlinear model is given by,
\begin{eqnarray}
K_{\rm NL} &=& 9\left(g_\omega\rho\frac{\partial\omega_0}{\partial\rho}
+ \frac{k_F^2}{3E_F^*} - g_\sigma\rho\frac{M^*}{E_F^*}\frac{\partial\sigma}{\partial\rho}
\right),\nonumber\\
\end{eqnarray}
with 
\begin{eqnarray}
\frac{\partial\sigma}{\partial\rho} =
\frac{a_1b_2+a_2b_3}{a_1b_1-a_3b_3}\quad\mbox{and}\quad 
\frac{\partial\omega_0}{\partial\rho} =
\frac{a_2b_1+a_3b_2}{a_1b_1-a_3b_3},
\end{eqnarray}
where
\begin{eqnarray}
a_1&=&m_\omega^2 + 3Cg_\omega^4\omega_0^2 + g_\sigma
g_\omega^2\sigma(2\alpha_1+\alpha_1'g_\sigma\sigma),\\
a_2 &=& g_\omega, \\
a_3 &=& -2g_\sigma g_\omega^2\omega_0(\alpha_1+\alpha_1'g_\sigma\sigma), \\
b_1 &=& m_\sigma^2 + 2A\sigma + 3B\sigma^2 - g_\sigma^2g_\omega^2\omega_0^2\alpha_1' 
\nonumber \\
&+&3g_\sigma^2\left(\frac{\rho_s}{M^*}-\frac{\rho}{E_F^*}\right), \\
b_2 &=& \frac{g_\sigma M^*}{E_F^*},\quad\mbox{and}\quad
b_3 = -a_3.
\end{eqnarray}
The Fermi momentum is $k_F$, and $E_F^*=(k_F^2+{M^*}^2)^{1/2}$. For the density dependent 
model, $K_{\rm DD}$ reads
\begin{eqnarray}
K_{\rm DD} &=& 9\left(\rho\frac{\partial\Sigma_R}{\partial\rho} 
+ \frac{2\Gamma_\omega\rho^2}{m_\omega^2}\frac{\partial\Gamma_\omega}{\partial\rho} 
+ \frac{\Gamma_\omega^2\rho}{m_\omega^2}
+ \frac{k_F^2}{3E_F^*}\right.
\nonumber\\
&+& \left.\frac{\rho M^*}{E_F^*}\frac{\partial M^*}{\partial\rho}\right),
\end{eqnarray}
with
\begin{eqnarray} 
\frac{\partial
M^*}{\partial\rho}&=&-\left(\Gamma_\sigma\frac{\partial\sigma}{\partial\rho}
+\sigma\frac{\partial\Gamma_\sigma}{\partial\rho}\right)\qquad\mbox{and}
\\
\frac{\partial\sigma}{\partial\rho}&=&\frac{\left[\rho_s-3\left(\frac{\rho_s}{M^*}
-\frac{\rho}{E_F^*}\right)\Gamma_\sigma\sigma\right]\frac{\partial\Gamma_\sigma}
{\partial\rho}+\frac{\Gamma_\sigma M^*}{E_F^*}}
{m_\sigma^2+3\left(\frac{\rho_s}{M^*}-\frac{\rho}{E_F^*}\right)\Gamma_\sigma^2},
\end{eqnarray}
observing the definitions of Eq.~(\ref{mfdd}). In the above expressions, $\rho$ and 
$\rho_s$ are obtained at $T=0$ regime by discarding in Eqs.~(\ref{rho})-(\ref{rhos}) the 
antiparticles distribution functions, and by replacing the particle distribution ones by 
the step function $\theta(k-k_F)$.

In  Ref.~\cite{rmft}, 128 Boguta-Bodmer parametrizations were analyzed
and the critical parameters showed an increasing behavior with $K_0$ (see Fig.~4).
Along that work, it was found that parametrizations 
with fixed values for the nucleon effective mass present $T_c$, $P_c$ and $\rho_c$ 
as clear functions of $K_o$ (see Fig.~5). 
\begin{figure}[!htb]
\centering
\includegraphics[scale=0.35]{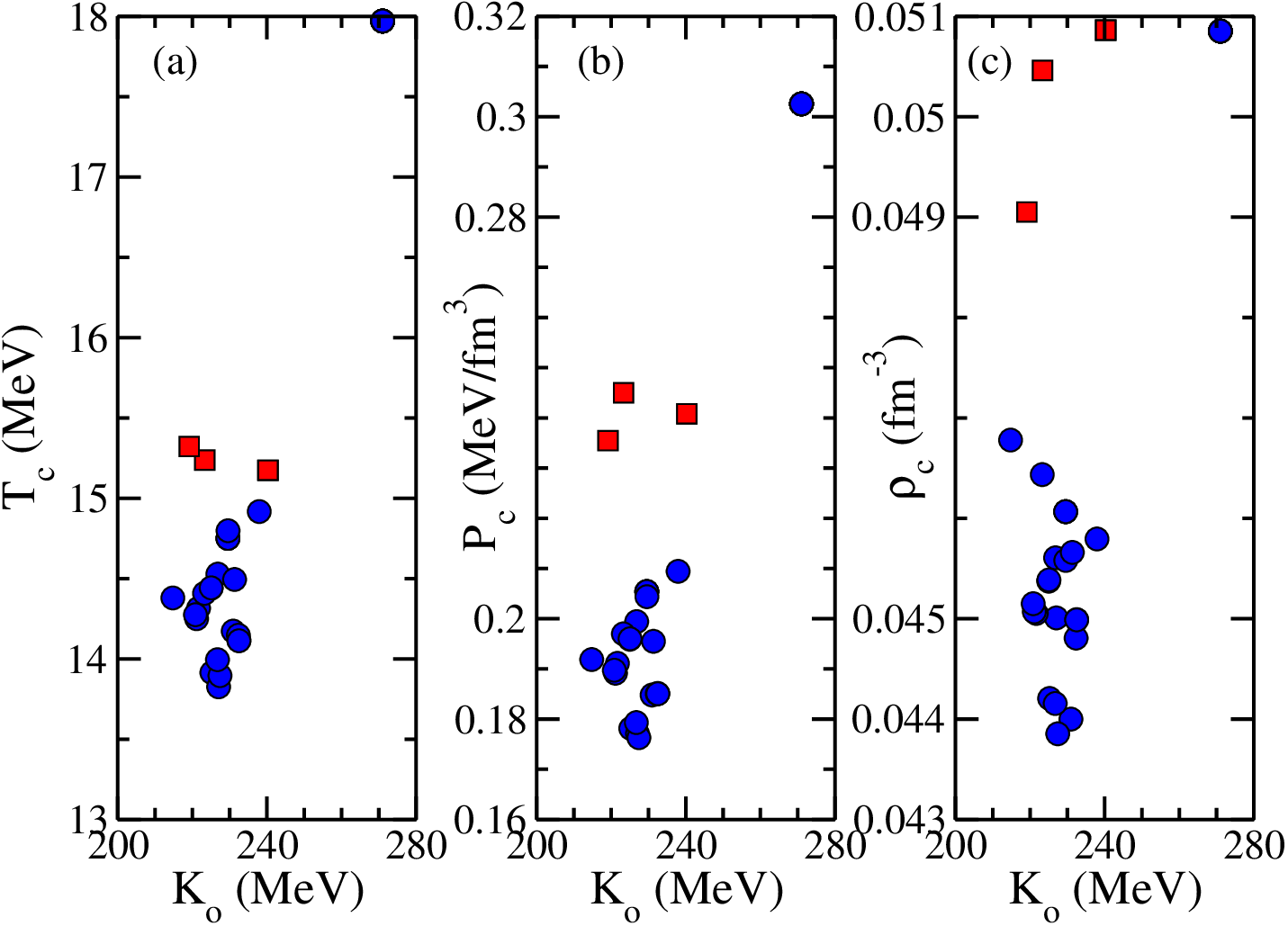}
\vspace{-0.2cm}
\caption{Critical (a) temperature, (b) pressure, and (c) density of CRMF 
parametrizations. Circles: nonlinear model. Squares: density dependent model.} 
\label{correlations}
\end{figure}

Here, we proceed in the same direction by displaying the critical parameters of the CRMF 
parametrizations as a function of $K_0$ in Fig.~\ref{correlations}. We separate the 
points concerning models with different structures, namely, the nonlinear model (circles) 
and the density dependent one (squares). If we consider only the model with more 
available data, i.e., the nonlinear one, we also verify an indication of $T_c$, $P_c$ and 
$\rho_c$ as increasing functions of $K_0$, as found in Ref.~\cite{rmft} for 
the less sophisticated Boguta-Bodmer model. Such general trends are also in line with 
recent results on classical models for real gases augmented with quantum statistical 
effects in the description of symmetric nuclear matter, see Ref.~\cite{vdw6}. In that 
work, the author provided the critical temperature of van der Waals, Redlich-Kwong-Soave, 
Peng-Robinson, and Clausius models. Models with higher values of $K_0$ also presented 
higher values of $T_c$ (see Fig.~3).

We have checked that there is no correlation between the critical
parameters and other important nuclear matter bulk properties, as the
symmetry energy and its slope.

\section{Summary and conclusions}

In the present work we have recalculated the critical parameters $T_c$, $P_c$ and 
$\rho_c$, which define the limiting point of the phase transition from
a gas to a liquid phase with 34 models, which have shown to satisfy
important nuclear matter constraints \cite{rmf} and reasonably
describe stellar matter macroscopic properties \cite{stars}. We have
divided these models into 6 categories and just two of them (Z271) and
(DD) approaches the experimental critical temperature values.
By comparing these observations with the neutron star main properties
calculated in Ref.~\cite{stars}, we see that only density dependent models
seem to behave well both at low and high densities, but this statement
requires a more consistent analyses and further experimental and observational
data. 

We have also verified that the critical parameters present a
correlation with the incompressibility, but the same is not true for 
other important nuclear matter bulk quantities, such as the energy
symmetry and its slope. 

Finally, we would like to mention that instabilities in
neutron-$\Lambda$ matter are also worth examining. The existence of
hypernuclei as bound systems \cite{hypernuclei} might imply that a
similar phase transition in an extended diagram with strangeness as an
extra degree of freedom is also present. In Ref.~\cite{vdw4}, spinodal
sections were obtained for matter with couplings fixed so that realistic
potentials were reproduced. Hence, investigations about the influence of strangeness on the
liguid-gas phase transition and the related critical points are under way.

\section*{Acknowledgments} 

This work was partially supported by Conselho Nacional de Desenvolvimento Cient\'ifico e 
Tecnol\'ogico (CNPq), Brazil under grants 300602/2009-0 and 306786/2014-1.

\end{document}